\documentclass[a4paper, 10pt]{article}
\newcommand{\iTitle}[1]{\begin{center}\Large\bf #1\end{center}}
\newcommand{\iAuthor}[1]{\begin{center}\small #1\end{center}}
\newcommand{\iAddress}[1]{\begin{center}\small\it #1\end{center}}
\newcommand{\iAbstract}[1]{\noindent {\large\bf Abstract}\\#1}
\newcommand{\iKeywords}[1]{\noindent{\it Keywords:} #1 }

\usepackage{multirow}
\usepackage{geometry}
\usepackage{indentfirst} 
\usepackage[pagewise,switch]{lineno}

\usepackage{amssymb}

\usepackage{graphicx}
\graphicspath{{figures/}}
\usepackage{subfigure}
\usepackage{pst-plot}
\usepackage{pst-blur}
\usepackage{pstricks}
\usepackage{pstricks-add}
\usepackage{color}
\usepackage{xcolor}
\definecolor{red}{rgb}{1,0,0}
\usepackage[colorlinks=true, linkcolor=blue, urlcolor=magenta, citecolor=red]{hyperref}
\newcommand{\refb}[1]{(\ref{#1})}
\newcommand{\iRef}[7]{#1, {\it``#2"}, \href{#7}{#3, {\bf#4}, #5(#6).} }

\newcommand{\bw}{\begin{widetext}}
\newcommand{\ew}{\end{widetext}}
\newcommand{\be}{\begin{equation}}
\newcommand{\en}{\end{equation}}
\newcommand{\bee}{\begin{equation}}
\newcommand{\ene}{\end{equation}}
\newcommand{\bea}{\begin{eqnarray}}
\newcommand{\ena}{\end{eqnarray}}
\newcommand{\eq}[1]{eq.~(\ref{#1})}
\newcommand{\eqref}[1]{Eq.~(\ref{#1})}

\def\pslash{p\!\!\!\slash }

\def\to{\rightarrow}

\def\pslash{p\!\!\!\slash }

\def\ie{{\it i.e.}}
\def\etc{{\it etc.}}

\begin{document}

\iTitle{Probing the noncommutative effects of phase space in the time-dependent Aharonov-Bohm effect}

\vspace{0.3cm}
\iAuthor{Kai Ma\footnote{\href{mailto:makainca@yeah.net}{makainca@yeah.net}}, Jian-Hua Wang}
\vspace{-0.6cm}
\iAddress{Department of Physics, Shaanxi University  of Technology, Hanzhong, 723001, Peoples Republic of China}

\iAuthor{Huan-Xiong Yang}
\vspace{-0.6cm}
\iAddress{Interdisciplinary Center for Theoretical Study, University of Science and Technology of China, Hefei 200026, Peoples Republic of China}

\vspace{0.5cm}
\noindent\rule[0.25\baselineskip]{\textwidth}{0.8pt}
\iAbstract{
We study the noncommutative corrections on the time-dependent Aharonov-Bohm effect when both the coordinate-coordinate and momentum-momentum noncommutativities are considered. This study is motivated by the recent observation that there is no net phase shift in the time-dependent AB effect on the ordinary space, and therefore tiny derivation from zero can indicate new physics. The vanishing of the time-dependent AB phase shift on the ordinary space is preserved by the gauge and Lorentz symmetries. However, on the noncomutative phase space, while the ordinary gauge symmetry can be kept by the Seiberg-Witten map, but the Lorentz symmetry is broken. Therefore nontrivial noncommutative corrections are expected. We find there are three kinds of noncommutative corrections in general: 1) $\xi$-dependent correction which comes from the noncommutativity among momentum operators; 2) momentum-dependent correction which is rooted in the nonlocal interactions in the noncommutative extended model; 3) momentum-independent correction which emerges become of the gauge invariant condition on the nonlocal interactions in the noncommutative model. We proposed two dimensionless quantities, which are based on the distributions of the measured phase shift with respect to the external magnetic field and to the cross section enclosed by the particle trajectory, to extract the noncommutative parameters. We find that stronger (weaker) magnetic field strength can give better bounds on the coordinate-coordinate (momentum-momentum) noncommutative parameter, and large parameter space region can be explored by the time-dependent AB effect.
}

\vspace{0.3cm}

\iKeywords{Noncommutative geometry, Aharonov-Bohm effect, Geometry phase}

\noindent\rule[0.25\baselineskip]{\textwidth}{0.8pt}

\tableofcontents
\newpage

\section{Introduction}\label{intro}
Since Seiberg and Witten pointed out that the space noncommutativity can appear from string theory embedded in a background magnetic field~\cite{SW:1999}, the physical effects of noncommutative space in various quantum systems have undergone extensive studies. On the noncommutative space, the coordinate operators subject to the following algebra,
\bee\label{eq:ncdefine}
[x_{\mu}, x_{\nu}] = i \theta_{\mu\nu}\,,
\ene
where $\theta_{\mu\nu}$ which has dimension of length-squared is a totally anti-symmetric constant tensor, and represents the strength and relative directions of the noncommutativity. The algebra \refb{eq:ncdefine} can be introduced by simple transformation 
\bee\label{eq:xshift}
x_{\mu} \to x_{\mu} + \frac{1}{2}\theta_{\mu\nu}p^{\nu} \,
\ene 
from the noncommutative space to commutative one~\cite{TCurtright:1998}. 
The commutation relation \refb{eq:ncdefine} indicates much richer nontrivial physics. The extraordinary constant vector, $\theta_{i}=\epsilon_{ijk}\theta_{jk}/2$, breaks the rotational symmetry~\cite{Douglas:2001,Szabo:2003}, and hence removes the degeneracy of the energy levels of the hydrogen atom~\cite{Chaichain:2001}. It also behaves like a magnetic dipole moment, and gives a contribution $\vec{\theta}\cdot\vec{B}$, and then affects the electromagnetic dynamics of the magnetic dipole moment. Furthermore, the motion of electric dipole momentum in the electromagnetic fields can also be influenced through the spin-orbital-like interaction, $\vec{\theta}\cdot(\vec{p}\times\vec{E})$~\cite{Zhang:2004,Ma:2011}. 

Particularly, the topological properties of the ordinary gauge field theory, such as the Aharonov-Bohm effect~\cite{AB:1959} and the Aharonov-Casher effect~\cite{AC:1984}, can also receive corrections~\cite{Chaichian:2001B,Chaichian:2002,Chaichian:2008,Li:2006,Ma:2016,Mirza:2004,Li:2007,Wang:2007,Mirza:2006,Li:2008}. The famous Aharonov-Bohm (AB) effect~\cite{AB:1959} predicts that for charged particle moving in a configuration with vanishing electromagnetic field, \ie~$F_{\mu\nu}(x)=0$, after a cyclic evolution the quantum state can acqurie a non-trivial phase shift proportional to the magnetic flux enclosed by the trajectory of the incident particle. It's importance is represented by the nonlocal interaction between the charged particle and the electromagnetic field $F_{\mu\nu}(x)$  in the quantum mechanics, which is formulated through the fundamental gauge potential $A_{\mu}(x)$ in the covariant derivative. Theoretically, two kinds of effect can be observed, and correspond to the vector potential $\vec{A}$ and the scalar potential $\varphi$, respectively. On the commutative space, the AB effect and its practical applications have been studied extensively. Most importantly, because the phase shift is measured by the fundamental magnetic flux $\Phi_{0}=h/e=4.13\times10^{-15}J\cdot s\cdot C^{-1}$ which is a very small quantity, the AB effect is also employed to probe the possible new physics at high energy scale~\cite{Chaichian:2001B,Chaichian:2002,Chaichian:2008,Li:2006}. 

However, it has been pointed out that the simple shift method in \eq{eq:xshift} can not lead to gauge invariant results~\cite{Chaichian:2008,Bertolami:2015}, and the nontrivial gauge invariant physical effects exist only for the noncommutative algebra in the momentum space as follows~\cite{Chaichian:2008,Bertolami:2015}
\bee\label{eq:npdefine}
[p_{\mu}, p_{\nu}] = i \xi_{\mu\nu}\,,
\ene
where $\xi_{\mu\nu}$ is also a totally anti-symmetric constant tensor, and parameterizing the momentum-momentum noncommutativity. By considering only the noncommutative relation \refb{eq:npdefine}, it was shown that the persistent charged current in a quantum ring pierced by magnetic flux can be sensitive to the spatial noncommtuativity~\cite{Liang:2015}. However, because the momentum operators are defined as the derivatives of the action with respect to the noncommutative coordinates, the algebra \refb{eq:npdefine} can appear naturally as a result of the configuration algebra \refb{eq:ncdefine}. Therefore, it is necessary to study the physical effects in the case that both the nontrivial algebra \refb{eq:ncdefine} and \refb{eq:npdefine} exists. On the other hand, it has been shown that the Seiberg-Witten (SW) map \cite{SW:1999} from the noncommmutative space to the ordinary one preserves the gauge symmetry, and has been proved to be very useful to investigate various problems on noncommutative space~\cite{Ma:2016,Martin:2012aw, Brace:2001, Barnich:2001, Picariello:2002, Banerjee:2002,Fidanza:2002,Jackiw:2002,SubirGhosh:2005,KayhanUlker:2008}. Therefore it is possible and pleasurable to study the physical effects of the nontrivial algebra \refb{eq:ncdefine} and \refb{eq:npdefine} by using the SW map.

In this paper, we study the noncommutative corrections on the time-dependent Aharonov-Bohm effect in the case of that both the coordinate and momentum algebras are nontrivial, \ie~both relations \refb{eq:ncdefine} and \refb{eq:npdefine} are taken into account. It has been shown that as long as the charging rate of the magnetic field is considerably larger than the typical scale of the particle traveling between the source and screen, there is no net phase shift in the time-dependent Aharonove-Bohm effects~\cite{Rousseaux:2008,Moulopoulos:2010,Singleton:2013,Singleton:2014,Singleton:2015,Mansoori:2016}. Therefore the time-dependent AB effect can be employed to investigate tiny derivation from zero in the phase shift which is induced by unknown new physics at high energy scale.

The contents of this paper are organized as follows: in Sec.~\ref{sec:nccovariantAB} we study the covariant formulation of the Aharonov-Bohm phase shift on the noncommutative phase space; in Sec.~\ref{sec:ncexpAB} we will study the noncommutative deformations on both time-independent and time-dependent Aharonov-Bohm phase shifts induced by the algebras \refb{eq:ncdefine} and \refb{eq:npdefine}; in Sec.~\ref{sec:exp}, we study the schemes to experimentally probing the phase space noncommutativity, and the experimental sensitivities on the noncommutative parameters $\theta_{\mu\nu}$ and $\xi_{\mu\nu}$; our conclusions are given in the final section, Sec. \ref{conclusion}.

\section{Covariant AB phase on noncommutative phase space}\label{sec:nccovariantAB}

The original AB effect is static in the sense that the external electromagnetic field is time-independent. Recently, it was shown that there is no net phase shift when the external electromagnetic field is time-dependent because the exact cancellation between the magnetic and electric contributions~\cite{Rousseaux:2008,Moulopoulos:2010,Singleton:2013,Singleton:2014,Singleton:2015,Mansoori:2016}. However, such cancellation can be broken on the noncommutative phase space because of the background field described by the noncommutative parameter $\theta_{\mu\nu}$ and $\xi_{\mu\nu}$. In this work, we will consider only the spatial noncommutativity, \ie, $\theta^{0i}=0$ and $\xi^{0j}=0$, because otherwise the unitary is broken. For the momentum-momentum noncommutativity \refb{eq:npdefine}, as we have mentioned, it can be introduced directly by using the shift method without destroying the original gauge symmetry~\cite{Chaichian:2008,Bertolami:2015},
\bee\label{eq:pshift}
p_{\mu} \to p_{\mu} - \frac{1}{2}\xi_{\mu\nu}p^{\nu}\,.
\ene
One of the nontrivial result of this transformation is that the commutator $[x_{\mu},\, p_{\nu}]$ is also deformed,
\bee\label{eq:xpcd}
\big[x_{\mu},\, p_{\nu}\big] 
= - i \bigg( \eta_{\mu\nu} - \frac{1}{4}\eta^{\alpha\beta} \theta_{\mu\alpha}\xi_{\beta\nu}\bigg)\,,
\ene
where $\eta^{\alpha\beta}=\rm{Diag}(1,-1,-1,-1)$ is the metric of the flat space-time. This means in one side the diagonal element and hence the Planck constant $\hbar$ is renormalized. In the other side it implies non-varnishing off-diagonal elements. Furthermore, the transformation \refb{eq:pshift} is not unique, and it has been proved that physical predictions, \ie~ expectation values, transition rates, \etc, are independent of the chosen transformation.
By considering only the transformation \refb{eq:pshift}, the Lagrangian for charged particle interacting with external electromagnetic fields can be written as
\bee\label{eq:lagrangian}
\mathcal{L} =
\bar{\psi}(x)( \pslash - Q ~\slash{\!\!\!\!A}_{NC} - m ) \psi(x) \,,
\ene
where $Q$ is the charge of the matter particle, and the total potential $A_{NC;\mu}$ is the sum of the original one $A_{\mu}$ and the effective term $A_{\xi;\mu}$ emerging from the transformation \refb{eq:pshift},
\bee\label{eq:Ashift}
A_{NC;\mu} = A_{\mu} + A_{\xi;\mu}\,,\,\,\;
A_{\xi;\mu} = \frac{\xi }{2Q} \big(0, -y,\, x,\, 0  \big)\,.
\ene
In general, the effect of the momentum-momentum noncommutativity is just a shift of the gauge potential which can contribute additional electromagnetic strength field $F_{\mu\nu}$. In the calculation of $A_{\xi;\mu}$, we have used the fact that the system that we are studying is two-dimensional, and therefore we have limited our analysis to the $x-y$ plane. The parameter $\xi$ corresponds to the component $\xi^{12}$.  

On the other hand, as we have mentioned, the coordinate-coordinate noncommutativity can not be introduced by directly using the shift method~\cite{Chaichian:2008,Bertolami:2015}. However, it has been shown that the Seiberg-Witten (SW) map \cite{SW:1999} from the noncommmutative space to the ordinary one preserves the gauge symmetry, and has been proved to be very useful to investigate various problems on noncommutative space~\cite{Ma:2016,Martin:2012aw, Brace:2001, Barnich:2001, Picariello:2002, Banerjee:2002,Fidanza:2002,Jackiw:2002,SubirGhosh:2005,KayhanUlker:2008}. Therefore we will use the SW map to study the noncommutative corrections induced by the nontrivial algebra \refb{eq:ncdefine}. For the $U(1)$ gauge symmetry the SW map is given by
\bea
\psi &\to & \psi - \frac{1}{2} Q \theta^{\alpha\beta} A_{NC;\alpha} \partial_{\beta}\psi\,,  \\
A_{NC;\mu} &\to & A_{NC;\mu} - \frac{1}{2} Q\theta^{\alpha\beta} A_{NC;\alpha} ( \partial_{\beta} A_{NC;\mu} + F_{NC;\beta\mu} )\,,
\ena
where $F_{NC;\mu\nu} = \partial_{\mu} A_{NC;\nu} -  \partial_{\nu} A_{NC;\mu} $ is the effective electromagnetic field strength tensor. Then in terms of the ordinary fields the noncommutative Lagrangian \refb{eq:lagrangian} can be written as,
\bee\label{ac-nc}
\mathcal{L}_{NC} =
\big( 1 - \frac{1}{4}Q\theta^{\alpha\beta} F_{NC;\alpha\beta} \big) \bar{\psi}(x) ( i\gamma_{\mu} D^{\mu}_{NC}  - m )  \psi(x) +
\frac{i}{2} Q\theta^{\alpha\beta} \bar{\psi}(x) \gamma^{\mu} F_{NC;\mu\alpha} D_{NC;\beta} \psi(x) \,,
\ene
where $D_{NC;\beta} =  \partial_{\beta} + i Q A_{NC;\beta}$ is the covariant derivative. This Lagrangian is gauge invariant because the noncommutative corrections depend only on the covariant derivative $D_{NC;\beta}$ and the electromagnetic field strength $F_{NC;\mu\nu}$. Therefore, the noncommutative corrections on the AB phase shift can be defined unambiguously, and can be interpreted consistently on the commutative and noncommutative phase spaces. The equation of motion can be obtained directly from this Lagrangian as follows,
\bee\label{ncmotioneq}
 ( i\gamma_{\mu} D^{\mu}_{NC}  - m )  \psi(x) +
\frac{i}{2} Q\big( 1 - \frac{1}{4}Q \theta^{\alpha\beta} F_{NC;\alpha\beta} \big) ^{-1} \theta^{\alpha\beta}\gamma^{\mu} F_{NC;\mu\alpha} D_{NC;\beta} \psi(x) =0\,,
\ene
where we have multiplied a factor $( 1 - Q \theta^{\alpha\beta} F_{NC;\alpha\beta}/4 )^{-1}$ to simplify the expression. By expanding $( 1 - Q \theta^{\alpha\beta} F_{NC;\alpha\beta}/4 ) ^{-1}$ with respect to the noncommutative parameter $\theta^{\mu\nu}$, one can see that the correction proportional to $Q\theta^{\alpha\beta} F_{NC;\alpha\beta}$ can be neglected at the first oder of the noncommutative parameter $\theta^{\mu\nu}$. In some studies this term was taken into account through the renormalization of the particle charge. However, for consistence, from here and after we will keep only the leading order terms. Therefore we will consider only the correction which involves the covariant derivative. Under this approximation, the equation of motion can be written as
\bea\label{ncacfinald}
&& ( i\gamma_{\mu} \mathcal{D}_{NC}^{\mu}  - m ) \psi(x) = 0 \,, 
\\[2mm]
\label{ncacfinald2}
&& \mathcal{D}_{NC}^{\mu}  = \big( \eta^{\mu}_{~~\beta} + \frac{1}{2} Q F_{NC}^{\mu\alpha}\theta_{\alpha\beta} \big) D^{\beta}_{NC} 
\equiv g^{\alpha}_{~~\beta}D^{\beta}_{NC} \,.
\ena
The deformed Dirac equation is similar to the Dirac equation in curved space-time with metric $g_{\alpha\beta}$. The leading order effect of the noncommutative phase space behaves like a perturbation on the flat space-time defined by the metric $\eta_{\alpha\beta}$. The perturbation depends on the external magnetic field, noncommutative parameters as well as their relative angles.

The AB phase can be obtained by solving the equation of motion, \ie~\eq{ncacfinald}. At the leading order of the noncommutative parameter, the AB phase shift can be written as
\bee
\phi_{NC}^{AB} = \phi^{AB} + \phi_{\xi}^{AB} + \phi_{\theta-v}^{AB} + \phi_{\theta-g}^{AB}\,.
\ene
The first term $\phi^{AB}$ is the ordinary AB phase~\cite{AB:1959}. The second term comes from the momentum-momentum noncommutativity. We have shown that the noncommutative correction coming from the nontrivial algebra \refb{eq:npdefine} only linearly shifts the vector potential, see \eq{eq:Ashift}. Therefore, the corresponding phase shift is
\bee
\phi_{\xi}^{AB} 
= \oint A_{\xi;\mu} d x^{\mu}
= \frac{1}{2}\int d S^{\mu\nu} F_{\xi;\mu\nu}
\ene
where $F_{\xi;\mu\nu} = \partial_{\mu}A_{\xi;\nu}- \partial_{\nu}A_{\xi;\mu}$. The third term $\phi_{\theta-v}^{AB}$ comes from the coupling between the perturbation on the metric and the matter field momentum, therefore it is the momentum-dependent noncommutative correction,
\bee\label{ncabp}
\phi_{\theta-v}^{AB} 
=  \frac{Q}{2} \oint  F^{\mu\alpha}_{NC} \theta_{\alpha\beta}  p^{\beta} d x_{\mu}
\,.
\ene
The third term $\phi_{\theta-g}^{AB}$ is the momentum-independent noncommutative correction, and comes from the deformation of the vector potential due to the perturbation on metric.
\bee\label{ncaba}
\phi_{\theta-g}^{AB} 
=  -\frac{Q^2}{2} \oint  F^{\mu\alpha}_{NC} \theta_{\alpha\beta}  A^{\beta}_{NC} d x_{\mu}
\,.
\ene
This term also ensures the gauge covariance of the noncommutative correction in \eq{ncacfinald2}. All the noncommutative corrections are gauge invariant and Lorentz covariant. Futhermore, unlike the case in the ordinary AB effect, both the momentum-dependent and momentum-independent phase shifts in \refb{ncabp} and \refb{ncaba} described by the local interactions between the charged particle and electromagnetic field strength $F^{\mu\nu}(x)$. 

\section{Noncommutative corrections on the time-dependent AB effect}\label{sec:ncexpAB}
To study the time-dependent AB effect, we employ the standard AB configuration, and consider an infinite solenoid with electric current $I(t)$ which creates a vector potential outside the solenoid as follows
\bee\label{afield}
\vec{A}(t, \vec{x}) = \frac{k I(t)}{r} \vec{e}_{\phi}\,,
\ene   
where $k$ is a constant whose exact form is not important for now and $\vec{e}_{\phi}$ is the unite vector in the azimuthal angle direction in the $x-y$ plane. We use the same gauge of vanishing electric potential as in Refs.~\cite{Rousseaux:2008,Moulopoulos:2010,Singleton:2013,Singleton:2014,Singleton:2015}, \ie, $\varphi=0$. In this case, the magnetic field and the electric field can be written as $\vec{B}(t) = \vec{\nabla}\times\vec{A}(t)$ and $\vec{E}(t) = - \partial_{t}\vec{A}(t)$, respectively. It has been shown that there is an exact cancellation between the magnetic and electric AB phase shifts, and therefore there is no net phase shift~\cite{Rousseaux:2008,Moulopoulos:2010,Singleton:2013,Singleton:2014,Singleton:2015}. However, because the noncommutative corrections directly involve the local interactions between the charged particle and the electromagnetic field strength, therefore the cancellation may not happen exactly. In this section we study the physical properties of the time-dependent AB effect on the noncommutative phase space.

Before we go on to study the time-dependent AB phase shift, it is necessary to discuss the time-independent AB phase shift in first. The first kind of correction comes from the momentum-momentum noncommutativity, and can be written as,
\bee
\phi_{\xi}^{AB} = \frac{\xi}{QB} \phi^{AB}\,,
\ene
where $\phi^{AB}$ is the ordinary AB phase shift. Because we consider only the spatial noncommutativity, \ie~$\xi^{0i}=0$ and also the local Lorentz invariance is broken, therefore this correction has no electric component. This property is important for the time-dependent AB effect because it can destroy the ordinary cancellation. We will discuss this later. On the other hand, a general property of both the momentum-dependent and momentum-independent noncommutative corrections is that the corrections involve the local interactions between the charged particle and electromagnetic field strength $F^{\mu\nu}(x)$, see \eqref{ncabp} and \eqref{ncaba}. However, for the momentum-dependent correction \refb{ncabp}, there is no singularity, and the electromagnetic field strength $F^{\mu\nu}(x)$ vanishes at the location of charged particle, therefore there is no net contribution. This property can also be understood by using the Stokes's theorem by which the momentum-dependent correction can be written as
\bee\label{ncabps}
\phi_{\theta-v}^{AB} 
=  \frac{Q}{4} \theta_{\alpha\beta}  p^{\beta}\oint  \partial^{\alpha}F^{\mu\nu}_{NC} d S_{\mu\nu}\,.
\ene
Because the electromagnetic field strength $F^{\mu\nu}$ is a constant over the whole space-time for time-independent AB effect, therefore $\partial^{\alpha}F^{\mu\nu} =0$ and hence there is no non-trivial phase shift. However, this is not true for the momentum-independent correction in \eqref{ncaba}. This is because there is a singularity in the vector potential $A_{\mu}$. 
By using the Stokes's theorem one can see that there is a non-vannishing term in the momentum-independent correction $\phi_{\theta-v}^{AB}$.  For the magnetic component we have,
\bee\label{ncabam}
\phi_{\theta-g}^{AB-M} 
=  -\frac{Q^2}{2} \oint  F^{ij}_{NC} \theta_{jk}  A_{NC}^{k} d x_{i}
 =  \frac{Q}{2} \big[\big( \vec{B} + \frac{1}{Q}\vec{\xi} ~\big)\cdot \vec{\theta}  ~\big]\phi^{AB} \,,
\ene
where we have used a relation $\vec{B}_{NC}\perp\vec{A}_{NC}$, \ie, $\vec{B}_{NC}\cdot\vec{A}_{NC}=0$, which is a result of the physical configuration \refb{afield}, and is necessary to create the singularity of the space. Therefore, there is a nontrivial contribution on the magnetic AB phase shift which is proportional to the magnetic flux passing through the fundamental area spanned by the noncommutative parameter $\vec{\theta}$. The electric AB phase shift is
\bee
\phi_{\theta-g}^{AB-E} 
=  -\frac{Q^2}{2} \oint  F^{0j}_{NC} \theta_{jk}  A^{k}_{NC} d t
= -\frac{Q^2}{2} \oint   \vec{\theta}\cdot (\vec{E}_{NC}\times\vec{A}_{NC}) d t
= 0 \,,
\ene
where we have used the static condition $\vec{E}_{NC} = \partial_{t}\vec{A}_{NC}=0$ for the time-independent AB phase. Therefore the net phase shift for the static AB effect can be written as
\bee\label{netncab}
\phi_{NC}^{AB} = \bigg[1 +  \frac{\xi}{QB} +\frac{1}{2} Q \bigg(\vec{B} + \frac{1}{Q}\vec{\xi} ~\bigg)\cdot \vec{\theta} ~\bigg]\phi^{AB}\,,
\ene
which scales the ordinary AB phase shift by a factor of $1 + \xi/QB + (Q \vec{B}\cdot \vec{\theta} + \vec{\xi} \cdot \vec{\theta} )/2$. In the consideration of that the noncommutative property of space happens at the Plank scale, and the large background from the ordinary AB effect, it is very hard to measure the spatial noncommutativity.

Let us go on to study the time-dependent AB phase shift on the noncommutative space. In this case the momentum-dependent correction can be non-zero, because the electromagnetic field does not vanish at the location of the charged particle wave. However, we will show that this kind of correction still vanishes because of the gauge invariance. For clarity, from here and after, we will add a tilde ``$~\widetilde{}~$" on the phase shifts in the time-dependent case to distinguish the time-independent case. The magnetic and electric phase shift of $\phi_{\theta-v}^{AB}$ are
\bee\label{eq:ncabpt}
\widetilde{\phi}_{\theta-v}^{AB-M} 
=  \frac{1}{2} Q m \theta_{\alpha\beta}v^{\beta}\oint  \partial^{\alpha} \vec{B}_{NC} \cdot d\vec{S}\,,
\ene
and
\bee
\widetilde{\phi}_{\theta-v}^{AB-E} 
= - \frac{1}{2} Q m \theta_{\alpha\beta}v^{\beta}\oint  \partial^{\alpha} \vec{A}_{NC} \cdot d\vec{x}
= - \widetilde{\phi}_{\theta-v}^{AB-M} \,.
\ene
In the derivation above, we have used the relation $\vec{E}_{NC}(t) = - \partial_{t}\vec{A}_{NC}(t)$. The results indicate an exact cancellation between the magnetic and electric phase shift in the momentum-dependent noncommutative correction. It is worthy to point out that this cancellation happens even in the case that there is time-position noncommutativity, \ie, $\theta^{0i}\neq0$.

For the momentum-independent noncommutative correction, because the derivation in \eqref{ncabam} is still valid for the time-dependent electromagnetic fields, therefore the form of the magnetic phase shift does not change in this case,
\bee\label{ncabamt}
\widetilde{\phi}_{\theta-g}^{AB-M}(t)
 =  \frac{1}{2} Q \bigg[\bigg(\vec{B} + \frac{1}{Q}\vec{\xi} ~\bigg)\cdot \vec{\theta} ~\bigg]\phi^{AB}(t) \,.
\ene
Here we have write down the time dependence explicitly.
For the electric phase shift, inserting the relation $\vec{E}_{NC}(t) = - \partial_{t}\vec{A}_{NC}(t)$ we have
\bee\label{ncabaet}
\widetilde{\phi}_{\theta-g}^{AB-E} 
= -\frac{Q^2}{2} \oint   \vec{\theta}\cdot (\vec{E}_{NC}\times\vec{A}_{NC}) d t
= \frac{Q^2}{2} \oint   \vec{\theta}\cdot (\partial_{t}\vec{A}_{NC}\times\vec{A}_{NC}) d t
= 0\,,
\ene
where we have used the general relation $\vec{A}_{NC}\times\vec{A}_{NC}=0$ in the second step. Therefore, for the momentum-independent noncommutative corrections on the time-dependent AB phase shift, there is no cancelation between the magnetic and electric components. Therefore the net time-dependent phase shift is,
\bee\label{ncabamt}
\widetilde{\phi}_{NC}^{AB}(t)
 =   \bigg[  \frac{\xi}{QB}  + \frac{1}{2} Q  \bigg(\vec{B}(t) + \frac{1}{Q}\vec{\xi} ~\bigg)\cdot \vec{\theta} ~\bigg]\phi^{AB}(t) \,.
\ene
Compared to the time-independent AB phase shift \refb{netncab}, there is no overwhelming background which exists on both commutative and noncommutative space for the time-independent Aharonov-Bohm effect. Therefore it is easier to measuring the spatial noncommutativity by using the time-dependent AB effect. It is well know that the interference pattern will have a shift of $N$ fringes when the phase shift equals $N$ times of the flux quanta, $\Phi_{0}=h/e=4.13\times10^{-15}J\cdot s\cdot C^{-1}$ which is a very small quantity. Therefore it is expected the time-dependent AB effect is sensitive to the noncommutative parameter. In addtion, the noncommutative correction is enlarged by the large magnetic flux $\phi^{AB}(t)$ over the surface enclosed by the orbit of charged particles.

\section{Detecting spatial noncommutativity}\label{sec:exp}
In last section we have obtained the noncommutative correction on the time-dependent AB phase shift. Comparing to the case that only the spatial noncommutativity is taken into account\cite{Ma:2016:TABncs}, there is an additional contribution that proportional to the dimensionless constant $\theta\xi$. On the other hand, in general the noncommutative correction \refb{ncabamt} oscillates with respect to time. Here we use the maximum value of $\widetilde{\phi}_{NC}^{AB}(t)$ to illustrate the method of probing the spatial noncommutativity, and estimate the experimental sensitivity to the noncommutative parameter. Furthermore in the following analysis we take $Q=1$ in unite of the fundamental electric charge $\big|e\big|$. 

The noncommutative parameters $\theta$ and $\xi$ can be extracted by using the distribution of the measured phase shift $\widetilde{\phi}_{NC}^{AB}$ with respected to the external magnetic field $B$, as well as to the effective cross section $S$ enclosed by the trajectory of the charged particle. In order to do this we define following two dimensionless quantities,
\bea
\alpha 
&\equiv& \frac{1}{B}\frac{d }{d S} \widetilde{\phi}_{NC}^{AB} 
= \frac{\xi }{B} + \frac{1}{2}\theta (B  + \xi)   \,,
\\[2mm]
\beta
&\equiv& \frac{1}{S}\frac{d }{d B} \widetilde{\phi}_{NC}^{AB} 
= \frac{1}{2}\theta (B  + \xi) \,.
\ena
The important property of $\alpha$ and $\beta$ is that it depends only on the external magnetic field. Therefore it is expected that the noncommutative parameter $\theta$ is proportional to the inverse of the external magnetic field, \ie~$\theta = c_{\theta} B^{-1}$, while the parameter $\xi$ is proportional to the external magnetic field, \ie~$\xi = c_{\xi} B$, where the coefficient $c_{\theta}$ and $c_{\xi}$ can be expressed in terms of $\alpha$ and $\beta$. The noncommutative parameters $\theta$ and $\xi$ can be obtained directly from $\alpha$ and $\beta$ by using the following inverse relations,
\bea
\xi &=& B(\alpha-\beta)\,,
\\[2mm]
\theta &=& \frac{1}{B}\frac{2\beta}{1+ \alpha-\beta}\,.
\ena
Therefore the coefficient can be written as follows,
\bea
c_{\theta} &=& \frac{2\beta}{1+ \alpha-\beta}\,,
\\[2mm]
c_{\xi} &=& \alpha-\beta\,.
\ena
If $\alpha=\beta$, then there is no momentum-momentum noncommutativity, and $c_{\theta}=2\beta$ in this limit. Factually the noncommutative parameter $\theta$ can be obtained directly through the second derivative of the measured phase shift with respect to the magnetic field,
\bee
\theta 
= \frac{1}{S}\frac{d^2 }{d^2 B} \widetilde{\phi}_{NC}^{AB} \,.
\ene
However measure the second derivative needs more experimental data, and it does not give further information, therefore we propose to use the quantities $\alpha$ and $\beta$.

To discuss the quantitative properties of the noncommutative corrections in \eq{ncabamt}, let us focus on the last term in \refb{ncabamt}, \ie,~the term proportional to the product of $\xi$ and $\theta$. Factually, this kind of correction also appears in the deformed coordinate-momentum commutator, see \eqref{eq:xpcd}, therefore it is constrained by the measurement on the Planck constant $\hbar$. Many experimental groups have precisely measured the Planck constant, and the level of the relative uncertainty is in the range of $10^{-6}$ to $10^{-9}$~\cite{2013:RichardSteiner}. In this paper we use the CODATA recommended value, $\hbar=1.054571800(13)\times10^{-34}J\cdot s$~\cite{2014:CODATA}, and hence we require
\bee\label{eq:planck}
|\theta||\xi| \le 4.92\times10^{-8}[\hbar^2]\,.
\ene 
This implies also that the momentum-momentum noncommutative parameter $\sqrt{\xi}$ is always smaller than the coordinate-coordinate noncommutatve parameter $\sqrt{|\theta|}^{-1}$ by a magnitude of $10^{4}$.

On the other hand, the Lorentz symmetry is also violated in the noncommutative models. The authors in Ref.~\cite{Carroll:2011} investigated the noncommutative effects of the algebra \refb{eq:ncdefine} in the clock-comparison test experiments, and obtained a lower bound of the space-space noncommutative parameter: $|\theta|^{-1/2} \ge 10\rm{TeV}$. This is a very strong constraint, but a little release is expected once the momentum-momentum noncommutativity is also taken into account. Here we employ a much more conservative point of view: the fundamental length scale $\sqrt{|\theta|}$ should be smaller than wave length of the particles under considering. In our case, measurement of the AB phase is usually performed for electron, which has a size of  about $1\rm{fm}$, and corresponds to a lower band $|\theta|^{-1/2} \gtrsim 1\rm{GeV}$. Therefore in the following studies, we will investigate the noncommutative parameter in the range $1\rm{GeV} \le |\theta|^{-1/2} \le 10\rm{TeV}$ (the upper bound is just used for truncating the range of parameter scanning). A similar conservative assumption is used in Ref.~\cite{Bertolami:2005}. The constraint \refb{eq:planck} implies $\xi \lesssim 10\rm{KeV}$ conservatively. By studying the deformation of the energy levels in the gravitational well, an upper band on the parameter $\xi$ was obtained: $|\xi|^{1/2} \lesssim 1\rm{meV}$~\cite{Bertolami:2005}. Under above constraints, and in the consideration that the experimental sensitivity of the phase shift is about $0.1\%$~\cite{ADCronin:2009}, as well as that the AB phase shift is at the oder of 1, \ie ~$\phi^{AB} \sim \mathcal{O}(1)$~\cite{Tonomura:1986}, we can concluded that the contribution proportional to the product of $\xi$ and $\theta$ is useless to put bounds on the noncommutative parameters $\xi$ and $\theta$.

For the contribution proportional to the noncommutative parameter $\xi$, it is essentially determined by the ratio of $\xi$ to the external magnetic field strength $B$. Therefor lower magnetic field can test larger parameter region of $\xi$. For an effective cross section of order $1\rm{cm}^2$  enclosed by the charged particle trajectory, in order to have a phase shift of order $1$, \ie~$\phi^{AB} \sim \Phi_{0}$, the magnetic field of order $10^{-10}\rm{T}$ is required. Taken this value as the lower limit of the experimental accessible magnetic field strength, and the experimental sensitivity of the phase shift as $0.1\%$, then the region
\bee
\sqrt{\xi} \gtrsim \rm{\mu eV},
\ene
can be investigated by using the time-dependent AB effect. This is $10^3$ times smaller then the strongest limit of parameter $\xi$ from the measurement on quantum gravitational states. Therefore vary large parameter space region of $\xi$ can be explored by using the time-dependent AB effect.

For the contribution proportional to the noncommutative parameter $\theta$, it is essentially determined by the ratio of the external magnetic field strength $B$ to the parameter $\theta^{-1}$. Therefore, experiment with stronger magnetic field strength can give better upper bound of $\theta^{-1}$. For a phase shift at order $1\rm{rad}$, and magnetic field strength $B=1\rm{T}$, then the noncommutative parameter $\theta^{-1}$ with region
\bee
\sqrt{\theta}^{-1} \lesssim 10\rm{GeV}\,
\ene
can be explored by using the time-dependent AB effect. Even through this region is below the strongest upper bound of $\sqrt{\theta}^{-1}$. However, this region can be enlarged by employing larger phase $\phi^{AB}$. For instance, if we still use an effective cross section $S=1\rm{cm}^2$, then the phase shift will be enlarged by a factor of about $10^{12}$. This means the region $\sqrt{\theta}^{-1} \lesssim\, 10^{3}\rm{TeV}$ can be explored. This is significantly higher then the current upper bounds on $\sqrt{\theta}^{-1}$. Therefore the time-dependent AB effect is also useful to probing the coordinate-coordinate noncommutativity.

\section{Conclusions}\label{conclusion}
In summary we studied the noncommutative corrections on the time-dependent Aharonov-Bohm effect when both the nontrivial coordinate-coordinate commutator \refb{eq:ncdefine} and the momentum-momentum commutator \refb{eq:npdefine} are considered. This study is motivated by the recent observation that there is no net phase shift in the time-dependent AB effect on the ordinary space~\cite{Rousseaux:2008,Moulopoulos:2010,Singleton:2013,Singleton:2014,Singleton:2015,Mansoori:2016}, and hence tiny derivation from zero can indicate new physics at high energy scale. The vanishing of the time-dependent AB phase shift on the ordinary space is because of the cancellation between the magnetic and electric contributions, that from the vector and scalar potential, respectively. Because both gauge and Lorentz transformations can interchange the roles of vector and scalar potential, therefore gauge and Lorentz symmetries are important in this phenomenon.

However, even through the gauge symmetry is preserved on the noncommutative space in order to get consistent and unambiguous physical results, the Lorentz symmetry are broken spontaneously by nonzero expectation values of the background fields which are described by the noncommutative parameters $\theta_{\mu\nu}$ and  $\xi_{\mu\nu}$. Therefore the exact cancellation between the magnetic and electric phase shifts can be destroyed by those background fields. In our study, we keep the ordinary gauge symmetry on the noncommutative phase space by using the SW map. Based on this method we studied both the time-independent and time-dependent AB effect. We find there are three kinds of noncommutative corrections in general: 1) $\xi$-dependent correction which comes from the noncommutativity among momentum operators; 2) momentum-dependent correction which is rooted in the nonlocal interactions in the noncommutative extended model; 3) momentum-independent correction which emerges become of the gauge invariant condition on the nonlocal interactions in the noncommutative model. 

For the $\xi$-dependent correction, even through the expression is similar to the ordinary phase, however because we considered only the spatial noncommutativity and hence magnetic phase shift is nonzero while the electric one vanishes. Therefore this kind of correction survive in both the time-independent and time-dependent AB effect. For the momentum-dependent noncommutative correction, there is a cancellation between the magnetic and electric phase shift for both the time-independent and time-dependent AB effect, just like the case on the commutative space. However, there is important contribution in the momentum-independent noncommutative correction. This is also true for both the time-independent and time-dependent Aharonov-Bohm effect. The advantage of the time-dependent Aharonov-Bohm effect is that there is no overwhelming background contribution coming from the commutative space. Therefore the time-dependent Aharonov-Bohm can be more sensitive to the spatial noncommutativity. 

To extract the noncommutative parameters $\theta$ and $\xi$, we proposed two dimensionless quantities, which are based on the distributions of the measured phase shift with respect to the external magnetic field and to the cross section enclosed by the particle trajectory. In this method, both $\theta$ and $\xi$ can be related to the external magnetic field by dimensionless coefficients $c_{\theta}$ and $c_{\xi}$ that are combination of $\alpha$ and $\beta$. The experimental sensitivities on the noncommutative parameters $\theta$ and $\xi$ are discussed by using the current experimental resolution of the phase shift which is about $0.1\%$~\cite{ADCronin:2009}. We find that stronger (weaker) magnetic field strength can give better bounds on the noncommutative parameter $\theta$ ($\xi$). We also shown that large parameter space region can be explored by the time-dependent AB effect. In conclusion we introduced a new approach to investigate the spatial phase space noncmmutativity. The advance of this approach is the clean background because of vanishing net time-dependent AB phase on the commutative space. The experimental schemes to probe the spatial noncommutativity, as well as the possible ways of improving the experimental sensitivityare also discussed.
\\[1cm]
\noindent\textbf{Acknowledgments}: 
K. M. is supported by the China Scholarship Council and the Hanjiang Scholar Project of Shaanxi University of Technology.  J. H. W. is supported by the National Natural Science Foundation of China under Grant No. 11147181 and the Scientific Research Project in Shaanxi Province under Grant No. 2009K01-54 and Grant No. 12JK0960. H.-X. Y. is supported in part by CNSF-10375052, the Startup Foundation of the University of Science and Technology of China and the Project of Knowledge Innovation Program (PKIP) of the Chinese Academy of Sciences.


\end{document}